\documentstyle[11pt,a4,cite,psfig,epsf]{article}
\newcommand{\lesssim}{ {\
\lower-1.2pt\vbox{\hbox{\rlap{$<$}\lower5pt\vbox{\hbox{$\sim$}}}}\ } }
\newcommand{\gtrsim}{ {\
\lower-1.2pt\vbox{\hbox{\rlap{$>$}\lower5pt\vbox{\hbox{$\sim$}}}}\ } }

\newcommand{\be}{\begin{equation}}
\newcommand{\ee}{\end{equation}}
\newcommand{\bea}{\begin{eqnarray}}
\newcommand{\eea}{\end{eqnarray}}
\newcommand{\noi}{\noindent}
\newcommand{\nn}{\nonumber}

\newcommand{\cL}{{\cal L}}
\newcommand{\cO}{{\cal O}}

\newcommand{\Imm}{\mbox{\rm Im}}

\newcommand{\tr}{\mbox{\rm tr}}

\newcommand{\GeV}{\mbox{\rm GeV}}

\newcommand{\annd}{\mbox{\rm and}}

\newcommand{\al}{\alpha}
\newcommand{\als}{\alpha_{\mbox{\rm {\scriptsize s}}}}

\newcommand{\Lmsb}{\Lambda_{\overline{\mbox{\rm\footnotesize MS}}}}

\newcommand{\GF}{G_{\mbox{\rm {\tiny F}}}}

\newcommand{\eff}{\mbox{\rm{\tiny eff}}}

\newcommand{\ra}{\rightarrow}

\newcommand{\hc}{\mbox{\rm h.c.}}

\newcommand{\h}{\mbox{\bf h}}

\input epsf

\begin{document}
\begin{titlepage}

\begin{flushright} CPT-98/P.3734\\ UAB-FT-460\\ 
\end{flushright}
\vspace*{1.5cm}
\begin{center}
{\Large \bf Matrix Elements of Electroweak Penguin Operators\\[0.2cm]
 in the
1/${N_c}$ Expansion}\\[3.0cm]
{\large {\bf Marc Knecht}$^{a}$,
{\bf Santiago Peris}$^b$ and {\bf Eduardo de Rafael}$^a$}\\[1cm]

$^a$  Centre  de Physique Th{\'e}orique\\
       CNRS-Luminy, Case 907\\
    F-13288 Marseille Cedex 9, France\\[0.5cm]
$^b$ Grup de F{\'\i}sica Te{\`o}rica and IFAE\\ Universitat Aut{\`o}noma de
Barcelona, E-08193 Barcelona, Spain.\\

\end{center}

\vspace*{1.0cm}

\begin{abstract}

It is shown that the $K\ra\pi\pi$ matrix elements of the four--quark operator
$Q_7$, generated by the electroweak
penguin--like diagrams of the Standard Model, can be calculated to first
non--trivial order in the chiral expansion and in the
$1/N_c$ expansion. Although the resulting $B$ factors $B_{7}^{(1/2)}$ and
$B_{7}^{(3/2)}$  are found to depend only logarithmically on the
matching scale $\mu$, their actual numerical values turn out to be  rather
sensitive to the precise choice of
$\mu$ in the $\GeV$ region. We compare our results to recent numerical 
evaluations from lattice--QCD and to other model estimates.
\end{abstract}

\end{titlepage}

\section{Introduction}
\label{sec:Introduction}

In the Standard Model, the physics of non--leptonic $K$--decays is
described by an
effective Lagrangian which is the sum of four--quark operators
$Q_i$ modulated by $c$--number coefficients $c_i$ (Wilson coefficients)
\be\label{eq:fourqL}
\cL_{\eff}^{\vert\Delta S\vert=1}=-\frac{\GF}{\sqrt{2}}V_{ud}V_{us}^{\ast}
\sum_{i}c_{i}(\mu)Q_{i}(\mu)+\hc \,,
\ee
where $\GF$ is the Fermi constant and $V_{ud}$, $V_{us}$ are the appropriate
matrix elements of flavour mixing. This is the effective Lagrangian which
results
after integrating out the fields in the Standard Model with heavy masses
$(Z^0,W^{\pm},t,b$ and $c$), in the presence of the strong
interactions evaluated in
perturbative QCD (pQCD) down to a scale
$\mu$ below the mass of the charm quark $M_{c}$. The scale $\mu$ has to be
large
enough for the pQCD evaluation of the
coefficients $c_i$ to be valid and, therefore, it is much larger than the
scale at
which an effective Lagrangian description in terms of the Nambu--Goldstone
degrees of freedom ($K$,
$\pi$ and $\eta$) of the  spontaneous $SU(3)_{L}\times SU(3)_{R}$ symmetry
breaking (S$\chi$SB) is appropriate. Furthermore, the evaluation of the
coupling constants of the low--energy effective chiral Lagrangian which
describes strangeness changing
$\vert\Delta S\vert=1$ transitions cannot be made within pQCD
because at scales
$\mu\lesssim 1\,\GeV$ we enter a regime where S$\chi$SB and confinement
take place
and the dynamics of QCD is then fully governed by non--perturbative phenomena.

In this letter we shall be concerned with two of the four--quark operators
in the
Lagrangian in eq.~(\ref{eq:fourqL}), the operators
\be
Q_7 = 6(\bar{s}_{L}\gamma^{\mu}d_{L})
\sum_{q=u,d,s} e_{q} (\bar{q}_{R}\gamma_{\mu}q_{R})\,,
\ee
and
\be
Q_8 = -12\sum_{q=u,d,s}e_{q}(\bar{s}_{L}q_{R})(\bar{q}_{R}d_{L})\,,
\ee
where $e_q$ denote quark charges in units of the electric charge,
and summation over
quark colour indices within brackets is understood. The operator $Q_7$
emerges at
the
$M_W$ scale from  considering  the so--called electroweak penguin
diagrams. In the presence of the strong interactions, the renormalization group
evolution of
$Q_7$ from the scale $M_W$ down to a scale $\mu\lesssim M_{c}$ mixes, in
particular, this operator
with the four--quark
density--density operator $Q_8$. These two operators, times their corresponding
Wilson coefficients, contribute to the lowest order $\cO(p^0)$
effective chiral Lagrangian which induces $\vert\Delta S\vert=1$ transitions
in the presence of electromagnetic interactions to order $\cO(\al)$ and
of virtual
$Z^{0}$ exchange, i.e., the Lagrangian~\cite{BW84}
\be\label{eq:order0}
\cL_{\chi,0}^{\vert\Delta S\vert=1}=-\frac{\GF}{\sqrt{2}}\frac{\al}{\pi}
V_{ud}V_{us}^{\ast}\,\frac{M_{\rho}^{6}}{16\pi^2}\,\h\,\tr
\left(U\lambda_{L}^{(23)}U^{\dagger}Q_{R}\right) +\hc \,.
\ee
Here, $U$ is the matrix field which collects the octet of pseudoscalar
Goldstone fields, $Q_{R}=\mbox{\rm diag}[2/3, -1/3, -1/3]$ is the right--handed
charge matrix associated with the electromagnetic couplings of the light
quarks, and
$\lambda_{L}^{(23)}$ is the effective left--handed flavour matrix
$\left(\lambda_{L}^{(23)}\right)_{ij}=\delta_{i2}\delta_{3j}$ $(i,j=1,2,3)$.
Under chiral  rotations ($V_{L},V_{R}$):
\be
U\ra V_{R}UV_{L}^{\dagger}\,,\quad Q_{R}\ra V_{R}Q_{R}V_{R}^{\dagger}\,,\quad
\lambda_{L}^{(23)}\ra V_{L}\lambda_{L}^{(23)}V_{L}^{\dagger}\,,
\ee
and the trace on the r.h.s. of eq.~(\ref{eq:order0}) is an invariant.
Actually, this is the only possible invariant which in the Standard Model
can generate
$\vert\Delta
S\vert=1$ transitions to orders
$\cO(\al)$ and $\cO(p^0)$ in the chiral expansion. 
With the given choice of the 
overall normalization factor
in front of the r.h.s. of eq.~(\ref{eq:order0}),
the coupling constant $\h$ is dimensionless and, {\it a priori}, of order
$\cO(N_c^{2})$ in the
$1/N_c$ expansion~\cite{'tH74,RV77,W79}. This coupling constant plays a 
crucial
r{\^o}le in the phenomenological analysis of radiative corrections to the
$K\ra\pi\pi$ amplitudes~\cite{deR88}. The determination of $\h$ is needed for
a reliable estimate of these corrections~\footnote{See ref. \cite{Ci98} for a recent discussion of the size of these corrections.}.  The purpose of
this note is to show that, following recent work reported in
ref.~\cite{KPdeR98}, one can calculate the
$K\ra\pi\pi$ matrix elements of the $Q_7$ operator (and, therefore, the so
called
$B_{7}^{(1/2)}$ and
$B_{7}^{(3/2)}$ factors) to first non--trivial order in the chiral expansion 
and the $1/N_c$ expansion~\footnote{A similar observation has also been made by J.~Donoghue~\cite{Do98}.}.
 This implies that (at least) the contribution to the  constant
$\h$ from the
$Q_7$ and $Q_8$ terms in the four--quark effective Lagrangian can be calculated
to first non--trivial order in the
$1/N_c$ expansion, a first step towards the required goal. The
leading $\cO(N_c^2)$ contribution to $\h$ vanishes trivially. It could only come
from the bosonization of the factorized
$Q_{8}$ operator times its Wilson coefficient, but this coefficient is
subleading at large $N_c$~\footnote{See e.g. Buras's
lectures~\cite{Bu98}.}. The
next--to--leading contribution comes from the bosonization of the unfactorized
$Q_7$ operator and its mixing via gluonic interactions with the $Q_{8}$
operator.
It involves the same two--point function which governs the electroweak $\pi^+
-\pi^0$ mass difference in large--$N_c$ QCD and this is why we are able to
compute the contributions from the $Q_7$ and $Q_8$ terms of the four--quark
Lagrangian to the constant
$\h$ at the stated order in the $1/N_c$ expansion.
The bosonization of the factorized $Q_7$ operator and of the unfactorized
$Q_8$ operator can only contribute to terms of order $\cO(p^2)$ (or higher) in
the chiral expansion and they are, therefore, inoperational in the calculation
of $\h$. It turns out, however, that
there is only a partial cancellation of the
$\mu$--dependences generated by the product of the bosonization
of the unfactorized
$Q_7$ operator times its Wilson coefficient $c_7$ with the $\mu$--dependence
coming from the product of the
Wilson coefficient $c_{8}$, which is non--leading in the $1/N_c$ expansion,
times the bosonization of the factorized
$Q_8$ operator. The full cancellation of $\mu$--dependences requires the
consideration, as well, of the bosonization of other four--quark operators (in
particular the unfactorized $Q_2$ operator) in the presence of the electroweak
interactions. This involves integrals of four--point functions which we
have not yet fully explored within the framework of the $1/N_c$ expansion.
The constant $\h$ gets, therefore, contributions from other operators than just
$Q_7$ and $Q_8$, and which we have not computed so far.
We shall, therefore, concentrate here on the calculation of the factors
$B_{7}^{(1/2)}$ and $B_{7}^{(3/2)}$, and on their comparison to recent lattice 
QCD determinations~\cite{Cetal98,Aetal98,Letal98},  as well as to recent 
analytic determinations which have been made~\footnote{See e.g. 
ref.~\cite{BFE98} and references therein.} using the ``effective action 
approach'' of ref.~\cite{PdeR91}.


\section{Bosonization of $Q_7$ and $Q_8$}
\label{sec:boson}

As already mentioned, the bosonization of the operator $Q_{7}$ is
needed to next--to--leading order in the $1/N_c$ expansion. The problem is
entirely analogous to the bosonization of the operator
$Q_{LR}\equiv\left(\bar{q}_{L}\gamma^{\mu}Q_{L}q_{L}\right)
\left(\bar{q}_{R}\gamma^{\mu}Q_{R}q_{R}\right)$ which governs the electroweak
$\pi^{+}-\pi^{0}$ mass difference and which has been recently discussed in
ref.~\cite{KPdeR98}. Because of the $LR$ structure, the factorized  component of
the operator $Q_{7}$, which is leading in $1/N_c$, cannot contribute to order
$\cO(p^0)$ in the low--energy effective Lagrangian. The first non--trivial
contribution from this operator is next--to--leading in the $1/N_c$
expansion and
is given by the integral~\footnote{Up to the replacement of $Q_L$ by 
$\lambda_{L}^{(23)}$, the bosonization of the operator $Q_7$ 
follows from the same procedure as described for the operator $Q_{LR}$ in 
ref.~\cite{KPdeR98}.}
\be\label{eq:LRgral}
Q_{7}\ra -3ig_{\mu\nu}\int \frac{d^4q}{(2\pi)^4}
\Pi_{LR}^{\mu\nu}(q)\,\,
\tr\left( U\lambda_{L}^{(23)} U^{\dagger} Q_{R}\right)\,,
\ee
involving the two--point function
\be\label{eq:lrtpf}
\Pi_{LR}^{\mu\nu}(q)=2i\int d^4 x\,e^{iq\cdot x}\langle 0\vert
\mbox{\rm T}\left(L^{\mu}(x)R^{\nu}(0)^{\dagger}
\right)\vert 0\rangle\,,
\ee
with currents
\be L^{\mu}=\bar{q}_{i}(x)\gamma^{\mu}\frac{1}{2}(1-\gamma_{5})q_{j}(x)
\quad \annd \quad
R^{\mu}=\bar{q}_{i}(x)\gamma^{\mu}\frac{1}{2}(1+\gamma_{5})q_{j}(x)\,,
\ee
and  $i$ and $j$ fixed flavour indices, $i\neq j$.
This integral, which has to be 
evaluated in the chiral limit, where ($Q^2= -q^2$)
\be\label{eq:lritpf}
\Pi_{LR}^{\mu\nu}(q)=(q^{\mu}q^{\nu}-g^{\mu\nu}q^2)\Pi_{LR}(Q^2)\,,
\ee
is divergent for large $Q^2$ and needs to be regulated. Before discussing this point, let us recall that in  
the large--$N_c$ limit, the spectral function associated with
$\Pi_{LR}(Q^2)$ consists of the difference of an infinite number of narrow
vector states and an infinite number of narrow axial--vector states,
together with
the Goldstone pion pole:
\be
\frac{1}{\pi}\Imm\Pi_{LR}(t) =\sum_{V}f_{V}^2 M_{V}^2\delta(t-M_{V}^2)
-\sum_{A}f_{A}^2 M_{A}^2\delta(t-M_{A}^2)-f_{\pi}^2\delta(t)\,.
\ee
Since $\Pi_{LR}(Q^2)$ obeys an unsubtracted dispersion relation, we find
that
\be\label{eq:LRN1}
-Q^2\Pi_{LR}(Q^2)=f_{\pi}^2+\sum_{A}f_{A}^2
M_{A}^2\frac{Q^2}{M_{A}^2+Q^2} -\sum_{V}f_{V}^2
M_{V}^2\frac{Q^2}{M_{V}^2+Q^2}\,.
\ee
Furthermore, in the chiral limit of QCD, the operator product expansion (OPE)
applied to the correlation function $\Pi_{LR}(Q^2)$ implies
\be
\lim_{Q^2\ra\infty} Q^2\Pi_{LR}(Q^2)\ra 0\,\quad\annd\quad
\lim_{Q^2\ra\infty} Q^4\Pi_{LR}(Q^2)\ra 0\,.
\ee
These relations result in the two Weinberg sum
rules~\cite{We67}
\be\label{eq:weinbergsrs}
\sum_{V}f_{V}^2 M_{V}^2-\sum_{A}f_{A}^2 M_{A}^2=f_{\pi}^2
\quad
\annd\quad
\sum_{V}f_{V}^2 M_{V}^4-\sum_{A}f_{A}^2 M_{A}^4=0\,.
\ee

The usual prescription \cite{bardeen} for the evaluation of integrals such as 
(\ref{eq:LRgral}) consists in taking a sharp cut-off in the (euclidian) 
integration over $Q^2$,
\be\label{eq:mugral}
Q_{7}\ra -6\frac{3}{32\pi^2}\int_{0}^{\Lambda^2} dQ^2
Q^2\left(-Q^{2}\Pi_{LR}(Q^2)\right)
\tr\left( U\lambda_{L}^{(23)} U^{\dagger} Q_{R}\right)\,.
\ee
With the constraints between the couplings and 
masses of the narrow states coming from the Weinberg sum 
rules~(\ref{eq:weinbergsrs}), the integral on the r.h.s. of eq.~(\ref{eq:LRgral})
becomes
then only logarithmically dependent on the ultraviolet scale $\Lambda$, with the
following result
\be\label{eq:explicit}
\int_{0}^{\Lambda^2} dQ^2 Q^2\left(-Q^{2}\Pi_{LR}(Q^2)\right)=
\left[\sum_{A}f_{A}^2 M_{A}^6
\log\frac{\Lambda^2}{M_{A}^2}-\sum_{V}f_{V}^2 M_{V}^6
\log\frac{\Lambda^2}{M_{V}^2}\right]\,.
\ee
Notice that if only the contribution from the Goldstone pole had been 
taken into account, the resulting expression would have displayed a 
polynomial dependence on the cut-off scale $\Lambda$. Another possibility is 
to evaluate the integral (\ref{eq:LRgral}) within a dimensional 
regularization scheme, say ${\overline{MS}}$, in which case one obtains the 
same result~(\ref{eq:explicit}), but with the correspondence between  
the cut--off $\Lambda$ and 
the ${\overline{MS}}$ subtraction scale $\mu$ given by  
\be\label{eq:scale}
\Lambda = \mu\cdot e^{\frac{1}{6}}\,.
\ee

On the other hand, at the level of approximation that we want to attain, the
bosonization of the $Q_{8}$ operator is only required to leading order in the
$1/N_c$ expansion, because its Wilson coefficient is already subleading.
To that order and to order $\cO(p^0)$ in the chiral expansion it
can be readily obtained from the bosonization of the factorized density
currents, with the result~\footnote{See e.g. the lectures in ref.~\cite{deR95}
and references therein.}
\be\label{eq:Q8bosonized}
Q_{8}\ra -12\left(2B\frac{f_{\pi}^2}{4}\right)^2
\tr\left( U\lambda_{L}^{(23)} U^{\dagger} Q_{R}\right)\,,
\ee
where $B$ is the low energy constant which describes the bilinear single 
flavour quark
condensate in the chiral limit, 
$B= - \langle{\bar\psi}\psi\rangle / f_{\pi}^{2}$ .

As discussed in ref.~\cite{KdeR97} there is a constraint that emerges in the
large--$N_c$ limit which relates the leading $d=6$ order parameter in the
OPE of the $\Pi_{LR}$ two--point function~\cite{SVZ79} to couplings and masses
of the narrow states:
\bea\label{eq:phi3}
\lim_{Q^2\ra\infty}Q^6\Pi_{LR}(Q^2) & = &
-4\pi^2\left(\frac{\alpha_s}{\pi}+\cO(\alpha_s^2)\right)
\langle\bar{\psi}\psi\rangle^2 \nonumber\\ & = & \sum_{V} f_{V}^2
M_{V}^6-\sum_{A} f_{A}^2 M_{A}^6\,.
\eea
This relation provides part of the cancellation between the $\mu$ dependence of
the bosonization of the operator $Q_{7}(\mu)$ with the short--distance
dependence on
$\mu$ in the Wilson coefficient of $Q_{8}$; but, as already mentioned, and
contrary to the simple case of the electroweak contribution to the $\pi^+
-\pi^0$ mass difference discussed in ref.~\cite{KPdeR98}, this cancellation
here is incomplete.
In this respect, we wish to comment on an important point concerning the scale
dependence in the relation in eq.~(\ref{eq:phi3}). The term on the r.h.s. of the
first line results from a lowest order pQCD calculation of the Wilson
coefficient. Using the renormalization group improvement to one loop, this
result
becomes
\be
-4\pi^2\frac{\alpha_s}{\pi}
\langle\bar{\psi}\psi\rangle^2\ra
-4\pi^2\langle\widehat{\bar{\psi}\psi}\rangle^2\frac{1}{-\beta_1}\left(
\frac{-\beta_{1}\als(Q^2)}{\pi}\right)^{\frac{2\gamma_{1}+\beta_{1}}{\beta_{1}}}
\,,
\ee
with $\langle\widehat{\bar{\psi}\psi}\rangle$ the scale invariant quark
condensate (the analog to invariant quark masses).  In the large--$N_c$ limit,
$\beta_{1}\ra
\frac{-11}{6}N_c$ and
$\gamma_{1}\ra \frac{3}{4}N_c.$ The $Q^2$ dependence of the one--loop result 
is indeed rather mild $\sim
\left(\log Q^2/\Lmsb^2\right)^{-2/11}$, but it does not go to a
constant as the exact large--$N_c$ result in the second line of
eq.~(\ref{eq:phi3}) demands. This mismatch is due to the fact that pQCD is at
best an approximation. It may happen that a two--loop renormalization group
improvement of the OPE result approaches a constant behaviour at large
$Q^2$ in a  better way. In any
case, this is a typical example of unavoidable mismatches that one will
encounter between non--perturbative evaluations of matrix elements and
short--distance  pQCD evaluations, which are necessarily only approximate. 
In general, however, it is well known that at higher orders ambiguities
will mix the short--distance coefficients of different powers in the OPE.
It is
difficult to imagine how to avoid these uncertainties in a final matching
between short--distances and long--distances unless a breakthrough is made in
understanding the relationship between pQCD and full QCD. 

Finally, if we restrict ourselves to the Lowest Meson 
Dominance (LMD) approximation to large-$N_c$ QCD discussed in 
ref.~\cite{PPdeR98} and identify $M_V$ with $M_\rho$, our 
calculation (in the ${\overline{MS}}$ scheme) gives the 
following contributions to the constant $\h$ coming from the 
operators $Q_7$ and $Q_8$,
\be
\frac{\al}{\pi}\h[Q_7]= -18c_7(\mu)\frac{f_\pi^2}{M_\rho^2}
\left[\log\frac{\Lambda^2}{M_\rho^2}-2\log 2\right]
\quad
\annd\quad
\frac{\al}{\pi}\h[Q_8] = -48\pi^2c_8(\mu)
\frac{\langle\bar{\psi}\psi\rangle^2(\mu)}{f_\pi^2 M_\rho^6}\ ,
\ee
where the relation (\ref{eq:scale}) is understood.

\section{The $B$ Factors $B_7^{(1/2)}$ and $B_7^{(3/2)}$}
\label{sec:bfactors}

The bosonic expression of $Q_7$ given by eqs.~(\ref{eq:mugral}) and
(\ref{eq:explicit}) enables us to compute the $K\to\pi\pi$
matrix elements induced by this operator which, following the usual conventions,
we express in terms of the following isospin amplitudes
\be
\langle Q_7\rangle_{I}\equiv \langle (\pi\pi)_{I}\vert Q_7 \vert K^{0}\rangle\,,
\qquad I=0,2\,.
\ee
To leading order $\cO (p^0)$ in the chiral expansion and to next--to--leading
order in the $1/N_c$ expansion, $\cO(1/\sqrt{N_c})$ for $K\ra \pi\pi$
amplitudes, we obtain the following result
\be\label{eq:Q7res}
\langle Q_7 \rangle_0
= \sqrt{2}\langle Q_7 \rangle_2
= \frac{6\sqrt{3}}{16\pi^2f_\pi^3}\,\left[\sum_{A}f_{A}^2 M_{A}^6
\log\frac{\Lambda^2}{M_{A}^2}-\sum_{V}f_{V}^2 M_{V}^6
\log\frac{\Lambda^2}{M_{V}^2}\right]\,.
\ee
It has become customary (rather unfortunately) to parameterize the results of
weak matrix elements of four--quark operators
$Q_i$ in terms of the factorized contributions from the
so--called vacuum saturation approximation (VSA), modulated by correction
factors $B_i^{(\Delta I)}$, $\Delta I =1/2, 3/2$. In the case of $Q_7$ and
$Q_8$, one then has
\be\label{eq:Q7VSA}
\langle Q_7 \rangle_0^{\rm VSA}
\,=\, \frac{1}{2}X + \frac{1}{2N_c}(Z+4Y) \, ,\ \
\langle Q_7 \rangle_2^{\rm VSA}
\,=\,-\frac{\sqrt{2}}{2}X+\frac{\sqrt{2}}{N_c}Y\,,
\ee
and
\be\label{eq:Q8VSA}
\langle Q_8 \rangle_0^{\rm VSA}
\,=\, \frac{1}{2}(Z+4Y) + \frac{1}{2N_c}X \, ,\
\langle Q_8 \rangle_2^{\rm VSA}
\,=\,\sqrt{2}Y-\frac{\sqrt{2}}{2N_c}X\, .
\ee
The quantities $X$, $Y$  and $Z$ which appear in the above expressions are the
same as those usually found in the literature, i.e.,
\bea
X &=& \sqrt{3}f_\pi(M_K^2-M_\pi^2)\,+\,\cO (p^4)
\, ,\nonumber\\
Y &=& \sqrt{3}f_\pi\left(\frac{M_K^2}{m_s+m_d}\right)^2\,+\,\cO (p^2)
\, ,\nonumber\\
Z &=& 4\left(\frac{F_K}{F_\pi}-1\right)Y\,+\,\cO (p^4)\, .
\eea
There is no theoretical justification to consider the VSA as a good limit of
any kind in QCD~\footnote{Obviously, it would have been much more reasonable to
normalize results to the large--$N_c$ result in the lowest order of the chiral
expansion.}.  This is reflected by
the fact that e.g., the terms proportional to $\frac{1}{N_c}$ in
eqs.~(\ref{eq:Q7VSA}) and (\ref{eq:Q8VSA}) do not correspond to the correct
$1/N_c$ expansion and, in fact, the contributions of
$Y$ and
$Z$ to
$\langle Q_7
\rangle_I^{\rm VSA}$, although suppressed by a factor $1/N_c$,  are actually
numerically dominant, since (throughout, $M_K$ denotes the neutral kaon mass)
\be\label{eq:YvsX}
Y \,=\, \frac{X}{M_K^2-M_\pi^2}\left(\frac{M_K^2}{m_s+m_d}\right)^2
\,\sim\,11.4 \, X \, \left(\frac{0.158\,{\rm GeV}}{m_s+m_d}\right)^2\, .
\ee
Furthermore, whereas $X$ is scale independent, $Y$ and $Z$ depend
on the ${\overline{MS}}$ subtraction scale $\mu$ through the quark mass term $m_s +m_d$ in the denominator.
We shall nevertheless follow these conventions, if only to be able
to compare our results to those existing in the literature.  The corresponding $B$ factors are then
defined as
\be
B_i^{(1/2)}\,=\, \frac{\langle Q_i \rangle_0}{\langle Q_i \rangle_0^{\rm VSA}}
\, ,\ \
B_i^{(3/2)}\,=\, \frac{\langle Q_i \rangle_2}{\langle Q_i \rangle_2^{\rm VSA}}
\, .
\ee
In the sequel, we quote our results in the ${\overline{MS}}$ scheme.

Considering first the operator $Q_7$, and  restricting the sums in
eq.~(\ref{eq:Q7res}) to the LMD approximation discussed in
ref.~\cite{PPdeR98}, the previous calculation  leads to the results
\be\label{eq:B7res0}
B_7^{(1/2)}(\mu )\,=\,
\frac{X}{X+\frac{1}{N_c}(Z+4Y)}\bigg\{
1\,+\,\frac{3}{2\pi^2f_\pi^2}\frac{M_V^4}{M_K^2-M_\pi^2}
\left[\log\frac{\Lambda^2}{M_V^2}-2\log 2\right]\bigg\}\, ,
\ee
and
\be\label{eq:B7res2}
B_7^{(3/2)}(\mu )\,=\,
\frac{X}{X-\frac{2}{N_c}Y}\bigg\{
1\,-\,\frac{3}{4\pi^2f_\pi^2}\frac{M_V^4}{M_K^2-M_\pi^2}
\left[\log\frac{\Lambda^2}{M_V^2}-2\log 2\right]\bigg\}\, .
\ee
If in these expressions we take the value $M_V =M_{\rho}$, we obtain, at the
scale $\mu=2$ GeV usually adopted in lattice calculations,
\be\label{eq:B7num}
B_7^{(1/2)}(\mu = 2\,{\mbox{\rm GeV}}) \,\sim \frac{1}{19.5}(1\,+\,23)\, ,\ \
B_7^{(3/2)}(\mu = 2\,{\mbox{\rm GeV}}) \,\sim -\frac{1}{6.6}(1\,-\,11.5)\, ,
\ee
for, say, $(m_s + m_d)(\mu = 2\,{\mbox{\rm GeV}}) = 158\,{\mbox{\rm MeV}}$, the
conventional reference normalization used in eq.~(\ref{eq:YvsX}). 
These values are both positive and greater than unity.
One should however notice that these numbers are rather sensitive
to the choice of the scale $\mu$ and/or to the value assigned to $M_V$~:
for instance, at $\mu=2M_V/e^{\frac{1}{6}}\sim 1.3$ GeV the 
contributions between square brackets on
the r.h.s.  of eqs.~(\ref{eq:B7res0}) and (\ref{eq:B7res2}) vanish
exactly. In order to illustrate
these uncertainties, we show in Fig.~1 the variation of 
these $B$ factors for a reasonable range of values of the scale $\mu$. 
The area between the solid lines 
corresponds to the choice $(m_s + m_d)(\mu = 2\,{\mbox{\rm GeV}}) = 
158\,{\mbox{\rm MeV}}$, with $\Lambda_{\overline{MS}}$ varied between 
300 MeV and 450 MeV. The area between the dashed lines reflects the same 
variation of $\Lambda_{\overline{MS}}$, 
but for the extreme low value 
$(m_s + m_d)(\mu = 2\,{\mbox{\rm GeV}}) = 100\,{\mbox{\rm MeV}}$ quoted in
some lattice results~\cite{gupta98}. 
Also shown is a magnification of the region corresponding to values 
of $\mu$ around 2 GeV, the reference scale at which the lattice 
results are usually given. We find that below 
$\mu\lesssim
1.3\,\GeV$ the $B_{7}^{(1/2)}$ and $B_{7}^{(3/2)}$ factors can even become
negative, a result which disagrees, drastically, with the positive values 
quoted at a ``matching scale'' of $0.8\,\GeV$  which are 
found in the constituent chiral quark model~\cite{BFE98}. It is, however, not
clear how this ``matching scale'' is related to the 
${\overline{MS}}$ scale, $\mu$, in QCD.

\vspace{0.4cm}
\centerline{\epsfbox{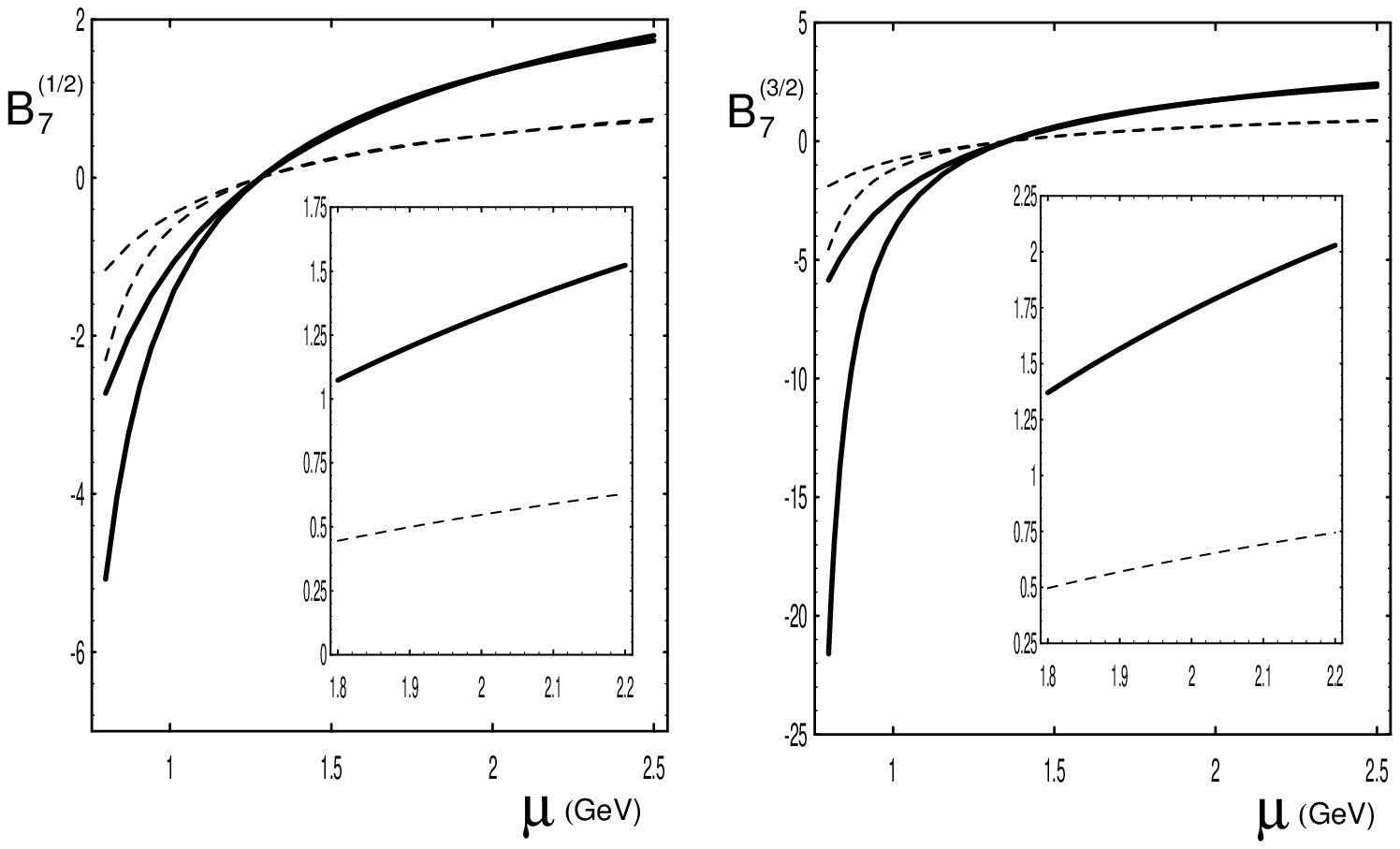}}
\vspace{0.2cm}

{\bf{Fig.~1}} {\it The $B$ factors, $B_{7}^{(1/2)}$ and $B_{7}^{(3/2)}$, as a
function of the
$\mu$ scale in $\GeV$. The solid lines correspond to 
$(m_s + m_d)(\mu = 2\,{\rm GeV}) = 158\,{\rm MeV}$; the dashed lines to the 
extreme low value $(m_s + m_d)(\mu = 2\,{\rm GeV}) = 100\,{\rm MeV}$.}

\vspace{0.3cm}

Several matrix elements of four--quark operators have also been obtained
in numerical simulations of lattice--QCD.
These numerical evaluations, however, are based on a yet different definition of
the $B$ parameters. The lattice definition uses a current algebra relation
between the
$K\to\pi\pi$ and the $K\to\pi$ matrix elements  which is in fact only valid at
order
$\cO (p^0)$ in the chiral expansion. Thus, in the case of $Q_7$ discussed here,
the lattice--QCD $B$ factor, which we shall distinguish with the symbol tilde
on top, is defined by the ratio
\be\label{eq:lattice}
{\widetilde B}_{7}^{(3/2)}\,\equiv\,\frac{\langle\pi^+
\vert Q_7^{(3/2)}\vert K^+\rangle}{\langle\pi^+\vert Q_7^{(3/2)}\vert
K^+\rangle^{\rm VSA}_0}\,,
\ee
where the matrix element in the denominator is evaluated in the chiral limit, as
indicated by the subscript ``0'', and the operator $Q_7$ has been decomposed
into its $\Delta I =1/2$ and  $\Delta I =3/2$ components, $Q_7 =
Q_7^{(1/2)}+Q_7^{(3/2)}$, with
\be
Q_7^{(3/2)}\,=\,2(\bar{s}_L\gamma^\mu d_L)\big[\bar{u}_R\gamma_\mu u_R -
\bar{d}_R\gamma_\mu d_R\big] \,+\,2(\bar{s}_L\gamma^\mu
u_L)(\bar{u}_R\gamma_\mu d_R)\,.
\ee

\noindent
The latest results obtained by various groups are consistent with each other:
\be\label{eq:B7lattice}
\left.
{\widetilde B}_{7}^{(3/2)}(\mu = 2\,{\mbox{\rm GeV}})\right\vert_{\rm latt} \,=\,
\left\{
\begin{array}{l}
0.58^{+5+2}_{-4-8}\\
0.61(11) 
\end{array}\right.
\ee
The first value has been taken from ref.~\cite{Letal98}, while the second one arises from the results of refs.~\cite{Cetal98} and \cite{Aetal98}, translated 
into the ${\overline{MS}}$ scheme~\cite{gimenez}.
In the vacuum saturation approximation which the lattice
community uses, one obtains
\be\label{eq:B7tildefact}
\langle\pi^+\vert Q_7^{(3/2)}\vert K^+\rangle^{\rm VSA}_0
\, =\,-\frac{2}{N_c}\frac{\langle\bar{\psi}\psi\rangle^2}{f_\pi^2}
\, =\,
-\frac{2}{N_c}\frac{f_\pi}{\sqrt{3}}Y\ ,
\ee
where in the second expression, following common practice, we have 
traded the dependence with respect to the condensate for the dependence 
on the strange quark mass, using the Gell-Mann--Oakes--Renner 
relation, which holds in the chiral limit. 
The calculation based on the bosonized
expression for this operator which we have discussed gives
\bea\label{eq:Q73/2result}
\langle\pi^+\vert Q_7^{(3/2)}\vert K^+\rangle & = &
\frac{3}{8\pi^2f_\pi^2}\,\left[\sum_{A}f_{A}^2 M_{A}^6
\log\frac{M_{A}^2}{\Lambda^2}-\sum_{V}f_{V}^2 M_{V}^6
\log\frac{M_{V}^2}{\Lambda^2}\right]\nn \\
 & \simeq & \frac{3}{4\pi^2}M_{V}^4\left(2\log
2-\log\frac{\Lambda^2}{M_{V}^2}\right)\,,
\eea
where in the second line we have used the LMD approximation discussed in
ref.~\cite{PPdeR98}. 
In this approximation  and with the same numerical input as
in (\ref{eq:B7num}), we obtain ${\widetilde B}_{7}^{(3/2)}(\mu = 2\,{\mbox{\rm
GeV}})\simeq +1.5$. However, as shown in Fig.~2, the
value of this $B$ factor is again very sensitive to the choice of the 
matching scale $\mu$. The results obtained for the smaller value of the 
strange quark mass ({\it i.e.} for larger values of the condensate)
are in agreement, within errors, with the numbers 
quoted in eq.~(\ref{eq:B7lattice}). Unfortunately, the values of $m_s$ (or of the condensate)
and of the $B$ factors obtained from the lattice are 
usually not quoted together.

In the case of $Q_8$, the large--$N_c$ limit of eq.~(\ref{eq:Q8VSA}) simply
reproduces the result obtained from the bosonized expression
(\ref{eq:Q8bosonized}). Since subleading corrections in the $1/N_c$ 
expansion of
this operator are not yet available, further comparison with lattice
 results for
$B_8^{(3/2)}$ has to be postponed.

\vspace{0.3cm}
\centerline{\epsfbox{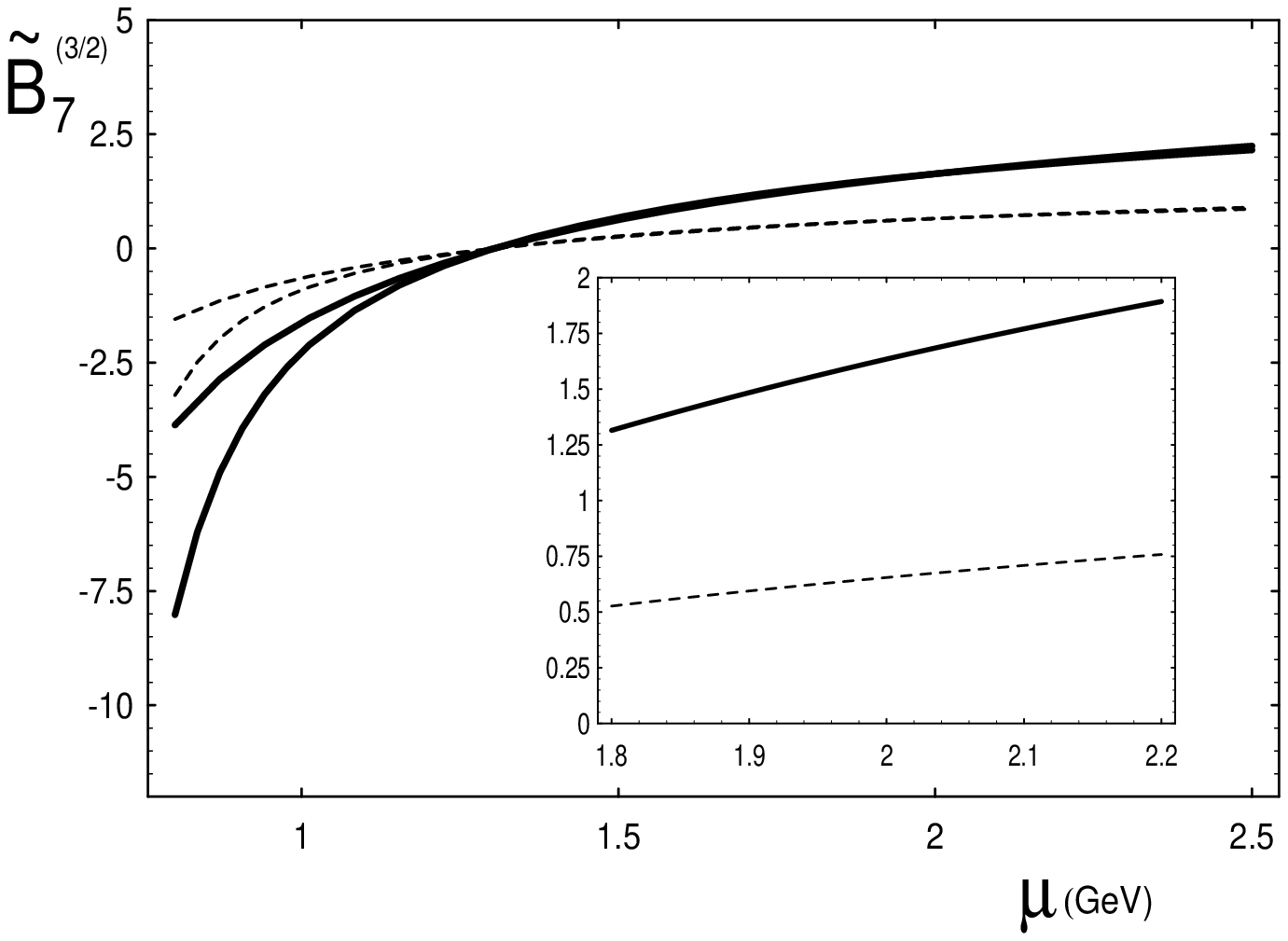}}
\vspace{0.3cm}

{\bf{Fig.~2}} {\it The $B$ factor, $\tilde{B}_{7}^{(3/2)}$ [see
eq.~(\ref{eq:lattice})], as a function of the
$\mu$ scale in $\GeV$. The solid lines correspond to
$(m_s + m_d)(\mu = 2\,{\rm GeV}) = 158\,{\rm MeV}$; the dashed lines to the 
extreme low value $(m_s + m_d)(\mu = 2\,{\rm GeV}) = 100\,{\rm MeV}$.}

\vspace{0.3cm}



\section{Conclusions and Outlook}
\label{sec:conout}

\noindent
The expressions  in eqs.~(\ref{eq:Q7res}) and (\ref{eq:Q73/2result}) are the
main results reported in this letter. They are a first step towards a systematic
evaluation of weak matrix elements in the chiral expansion and to first
non--trivial order in the $1/N_c$ expansion. Our final aim, however, is to
obtain values for the coupling constants of the low energy $\Delta S=1$ and
$\Delta S=2$ chiral effective Lagrangian directly; i.e., constants like $\h$ in
eq.~(\ref{eq:order0}) and not of individual matrix elements of four--quark
operators.  It is encouraging from the results obtained so far, to find such
simple analytic expressions which exhibit only a logarithmic dependence on the
matching scale
$\mu$;  however, the fact that the numerical results are so sensitive to
the choice of
$\mu$ in the $\GeV$ region is perhaps an indication that one should be extremely
cautious in the evaluation of errors of
$B$ factors, in general, both in model calculations and in lattice QCD
numerical simulations.

\vspace*{7mm}
{\large{\bf Acknowledgments}}

\vspace*{3mm}

\noi
We wish to thank M.~Perrottet and L. Lellouch for
discussions on topics related to the work reported here, V. Gimenez for 
an informative correspondance, and 
the referee for constructive remarks.  This work has been
supported in part by TMR, EC-Contract No. ERBFMRX-CT980169 (EURODA$\phi$NE). 
The work of S.~Peris has also been partially supported by the research 
project CICYT-AEN98-1093.







\begin{thebibliography}{99}

\bibitem{BW84}
         J.~Bijnens and M.~Wise, Phys. Lett. {\bf 137B} (1984) 245.

\bibitem{'tH74}
         G.~'t Hooft, Nucl. Phys. {\bf B72} (1974) 461;
         {\it ibid.} Nucl. Phys. {\bf B73} (1974) 461.

\bibitem{RV77}
         G.~Rossi and G.~Veneziano, Nucl. Phys. {\bf B123} (1977) 507.

\bibitem{W79}
         E.~Witten, Nucl. Phys. {\bf B160} (1979) 57.

\bibitem{deR88}
        E.~de Rafael, Nucl. Phys. {\bf B} (Proc. Suppl.) {\bf 7A} (1989)

\bibitem{Ci98}
        V.~Cirigliano, J.F.~Donoghue and E.~Golowich, hep-ph/9810488.
         
\bibitem{KPdeR98}
         M.~Knecht, S.~Peris and E.~de Rafael, Phys. Lett. {\bf 443B} (1998) 
         255.

\bibitem{Do98}
         J.~Donoghue, Communication at the Bad--Honnef Workshop on Chiral
         Effective Theories, 30 Nov.--4 Dec. 1998 (unpublished).


\bibitem{Bu98}
         A.J.~Buras, ``Weak Hamiltonian, CP Violation and Rare Decays'',
               Les Houches lectures 1997, hep-ph/9806471.

\bibitem{Letal98}
         L.~Lellouch and C.-J.~David Lin, hep-lat/9809142.

\bibitem{Cetal98}
         L.~Conti {\it et al.}, Phys. Lett. {\bf B421} (1998) 273.

\bibitem{Aetal98}
         C.R.~Allton {\it et al.}, hep-lat/9806016.

\bibitem{BFE98}
         S.~Bertolini, M.~Fabbrichesi and  J.O.~Eeg, hep-ph/9802405.

\bibitem{PdeR91}
         A.~Pich and E.~de Rafael, Nucl. Phys. {\bf B358} (1991) 311.

\bibitem{We67}
        S.~Weinberg, Phys. Rev. Lett. {\bf 18} (1967) 507.

\bibitem{bardeen}
W.A.~Bardeen, A.J.~Buras and J.-M.~G{\'e}rard, Nucl. Phys. {\bf B293}
        (1987) 787; Phys. Lett. {\bf B192} (1987) 138; {\bf B211} (1988) 
        343.\\
        W.A.~Bardeen, J.~Bijnens and J.-M.~G{\'e}rard, Phys. Rev. Lett. 
        {\bf 62} (1989) 1343.\\
        For a review and further references, see A.J.~Buras,
        ``The 1/N Approach to Non-leptonic
        Weak Interactions'', in {\it CP Violation}, ed. C. Jarlskog
        (World Scientific, Singapore, 1989).

\bibitem{deR95}
         E.~de Rafael, ``Chiral Lagrangians and Kaon CP--Violation'', in
         {\it CP Violation and the Limits of the Standard Model}, Proc.
         TASI'94, ed. J.F.~Donoghue (World Scientific, Singapore, 1995).

\bibitem{KdeR97}
         M.~Knecht and E.~de Rafael, Phys. Lett. {\bf B424} (1998) 335.

\bibitem{SVZ79}
        M.A.~Shifman, A.I.~Vainshtein and V.I.~Zakharov, Nucl. Phys. {\bf
        B147} (1979) 385, 447.

%
%

\bibitem{PPdeR98}
         S.~Peris, M.~Perrottet and E.~de Rafael, JHEP {\bf 05}
         (1998) 011.

\bibitem{gupta98}
         R.~Gupta, `` Introduction to Lattice QCD'', Les Houches lectures 1997,         hep-lat/9807028, and references therein.

\bibitem{gimenez}
         V. Gimenez, private communication.

\end{thebibliography}
\end{document}